\documentclass[hyper, notoc]{JHEP3}
\usepackage{amsmath,amssymb,amsfonts,epsfig}  

\newcommand{\be}{\begin{equation}}
\newcommand{\ee}{\end{equation}}
\newcommand{\bea}{\begin{eqnarray}}   
\newcommand{\eea}{\end{eqnarray}}

\title{The Higgs Mass Bound in Gauge Extensions of the Minimal Supersymmetric Standard Model}
\author{Puneet Batra\thanks{Present Address: Department of Physics and Astronomy, Johns Hopkins University, 
Baltimore, MD 21218} \\ Department of Physics, Stanford University, Stanford, CA 94305}
\author{Antonio Delgado, David E. Kaplan\\Department of Physics and Astronomy, Johns Hopkins University, 
Baltimore, MD 21218}
\author{Tim M.P. Tait \\ Fermi National Accelerator Laboratory, Batavia, IL 60510}

\abstract{
The minimal supersymmetric standard model, and extensions, have stringent upper bounds
on the mass of the lightest Higgs boson if perturbativity up to the Planck scale is assumed.
We argue that these bounds are softened tremendously if the Higgs is charged under an
asymptotically free gauge group.
We present a model with an additional $SU(2)$ gauge group
which easily produces Higgs masses above 200 GeV while avoiding electroweak constraints.  
If one allows some fine-tuning
of the high-scale value of the gauge coupling, Higgs masses greater than 350 GeV are achieved.
Unification of couplings is predicted to similar accuracy as in the minimal supersymmetric standard model 
with only small deviations at the two-loop level.}

\preprint{FERMILAB-PUB-03/245-T \\ SU-ITP-03/25-T}
\begin{document}

\section{Introduction}
The minimal supersymmetric standard model (MSSM) is perhaps the best motivated
example of new physics at the weak-scale.  Among its virtues
are its elegant explanation of the stabilization of the electroweak scale, 
its dramatic prediction of gauge coupling unification and its low impact 
on electroweak precision measurements.  In fact, the MSSM prediction 
of a light Higgs is favored by current data \cite{Abbaneo:2001ix}.

However, the MSSM is getting squeezed.  The lightest Higgs is, at 
tree-level, lighter than the $Z$ boson.  This mass range has been
excluded by LEP-II, and would rule out the MSSM if not for the fact that large
quantum corrections from the top sector can raise the
Higgs mass to 130 GeV---though only in the case of large $\tan\beta$
(the ratio of the VEV's of the two Higgses), 
1 TeV stop masses, and a maximal stop mixing angle \cite{Carena:1995wu}. 
The tension in the MSSM comes from the dual role played by the stops.  
On the one hand, the stops cut off the quadratic divergence of the top loops
and thus should be no heavier than the electroweak scale to avoid fine-tuning.
On the other hand, they must be heavy enough to generate a Higgs mass above
the LEP-II bound.

The current experimental situation naturally leads one to consider
extensions of the MSSM which relieve this tension.  In particular, the
tree-level bound on the Higgs mass is a direct consequence of the strength
of the Higgs quartic coupling. In the MSSM, the Higgs quartic coupling
comes only from the $D$-terms of the electroweak gauge groups and is
therefore a fixed function of the (relatively small) electroweak gauge
couplings.

The physical Higgs mass can be increased by
enhancing the quartic coupling through extended gauge sector and/or new superpotential Higgs
couplings \cite{Haber:1986gz}.  
This has been widely studied in the context of
extra singlet or triplet Higgses \cite{Espinosa:1998re}.  For singlet and
triplet soft masses of 1 TeV the bound on
the Higgs mass is found to be approximately 150 GeV and 200 GeV respectively. A 
thorough analysis of the electroweak precision and fine-tuning constraints in both types of 
models would be worthwhile. 

Such models are limited in the bound on the Higgs mass by requiring perturbativity of all 
couplings up to the GUT scale.  The issue:  {\it every contribution to the Higgs quartic coupling is infrared free}.
Couplings run from an already perturbative value at the GUT scale to smaller values near the weak scale.  
Thus we argue that in order to significantly increase the bound, the Higgs must be charged under an 
asymptotically-free group.  This will require extending the gauge group of the MSSM.

The additional contribution to the quartic potential here is the $D$-term of the asymptotically free group.
When the full gauge group breaks down to that of the standard model just above the
weak scale, this new $D$-term would quickly decouple in the supersymmetric limit.
Therefore, the field responsible for breaking the gauge symmetry must have a 
supersymmetry-breaking mass at or above the breaking scale.  The effect is to retain
the $D$-term in the potential at low energies.
To avoid large unwanted contributions to electroweak precision
measurements from the new gauge bosons, the breaking scale, and therefore the soft mass,
should be in the multi-TeV range.  

In our scenario, supersymmetry breaking can be much larger than the 
weak scale while retaining naturalness due to the (amazing) properties of $D$-terms:
\begin{itemize}
\item{After integrating out the field which breaks the gauge symmetry 
at the multi-TeV scale, we have an effective 
hard breaking of supersymmetry in the quartic
sector of the model while the gauge and top sectors are still supersymmetric
(broken softly).  There is no one-loop quadratically divergent contribution to 
the Higgs mass parameter from the top sector cut off by this higher scale, but there is
one proportional to the {\it additional} quartic term.}
\item{While breaking a gauge symmetry in a non-supersymmetric 
way would in principle produce both quartic and mass terms for the 
Higgs fields from the $D$-term, a VEV in a $D$-flat direction
leaves all fields without VEVs at that scale ({\it e.g.}, the Higgs) 
massless at tree-level.}
\item{The multi-TeV soft mass for the breaking field feeds into the 
Higgs mass renormalization group equations only at two loops and 
does not destabilize the weak scale.  In addition, the breaking field 
quantum numbers typically disallow any renormalizable superpotential 
couplings to MSSM fields.}
\end{itemize}

\section{Warm-up: $U(1)_x$}

Before we present our model we present a warm-up version with an extra
$U(1)$ in which the non-decoupling of the $D$-term is in effect, though the group is
of course non-asymptotically free.  We then present a model with an asymptotically-free
$SU(2)$.  The dominant  constraint on this model comes from electroweak precision
measurements and the desire for naturalness in couplings.

Now take the MSSM and gauge a $U(1)_x$.  A simple choice of charges is
the $\tau^3$ generator of $SU(2)_R$, namely $Q,U^c,D^c,L,E^c$ have charges
$0,-\frac{1}{2},+\frac{1}{2},0,+\frac{1}{2}$ respectively, $\overline{H}, H$ 
have charges 
$\pm \frac{1}{2}$ and three generations
of right-handed neutrinos $N^c$ are added with charges $-\frac{1}{2}$ to 
cancel the $U(1)_x^3$ anomaly.

In addition to the MSSM $D$-terms,
\begin{equation}
\frac{g^2}{8} \left( 
H^\dagger\sigma^a H -\overline{H} \sigma^a \overline{H}^{\dagger} + 
\ldots \, \right)^2
+\frac{g_Y^2}{8}\! \left( |\overline{H}|^2 - |H|^2+\ldots \, \right)^2,
\end{equation}
we have an additional contribution to the Higgs
potential coming from the new $U(1)_x$ $D$-term:
\begin{equation}
\frac{g^2_x}{2} \left[ \frac{1}{2} \left| \overline{H} \right|^2 - \frac{1}{2} 
\left| H \right|^2
+ q \left| \phi \right|^2 - q \left| \phi^c \right|^2 + \ldots \right]^2
\end{equation}
where the ellipsis represents the rest of the charged MSSM scalars. 
The $\phi,\phi^c$
fields, responsible for breaking $U(1)_x$, are uncharged under the 
MSSM gauge group and have charges $\pm q$ under $U(1)_x$.  Their scalar 
potential comes from
the superpotential ${\mathcal W}=\lambda S (\Phi \Phi^c - w^2)$
and a soft mass $m_\phi^2$, giving a potential
\begin{equation}
V_{\phi}=\lambda^2 \left|\phi\right|^2 \left|\phi^c\right|^2 - B \phi \phi^c 
+ h.c.
+ m_{\phi}^2 \left(\left|\phi\right|^2 + \left|\phi^c\right|^2\right) \ ,
\end{equation}
where $B \equiv \lambda w^2$ and all couplings are made real and positive 
from field redefinitions (except $m_\phi^2$ which is automatically real and
 taken to be positive). We assume $\phi$ and $\phi^c$ soft masses are 
the same due to ultraviolet dynamics.

For $B> m_\phi^2$, we have
$\langle\phi\rangle^2 = \langle\phi^c\rangle^2 = (B-m_\phi^2) / \lambda^2$.
Taking $B\gg v^2$, we 
integrate out the $\phi$ 
fields at tree-level and find an extra contribution to the MSSM Higgs potential:
\begin{eqnarray}
\frac{g_x^2}{2}\left(\frac{1}{2} \left| \overline{H} \right|^2 - \frac{1}{2} 
\left| H \right|^2\right)^2
	&\times& \left(1+\frac{M_{Z^\prime}^2}{2 m_\phi^2} \right)^{-1},
\end{eqnarray}
where $M_{Z'}=2qg_x \langle \phi \rangle$. In order to maximize the
contribution to the Higgs mass, we would like a large but perturbative
$g_x$ and a small $U(1)_x$ gauge boson mass compared to $m_\phi$.  At
the same time we don't want a soft mass so large that it destabilizes the
weak scale.

Electroweak symmetry breaking occurs under the same conditions as in the MSSM.
The adjusted tree-level bound for the CP-even Higgs mass is
\be
\label{tree-higgs}
m^2_{h^0} < \left(\frac{g^2}{2}+ \frac{g'^2}{2} 
+ \frac{g_x^2}{2} \left(1+\frac{M_{Z^\prime}^2}{2 m_\phi^2} \right)^{-1}\right)
v^2 \cos^2{2\beta},
\ee
where the inequality is saturated in the ``decoupling limit'' when the CP-odd
Higgs mass $m_{A}^2$ is much larger than $m_{h^o}^2$.

Electroweak precision measurements put a lower limit on the
$Z^\prime$ mass as a function of its couplings.  There are oblique
corrections at order $v^2 / \langle \phi \rangle^2$ to the $Z$
boson mass through its mixing with the $Z^\prime$, and non-oblique
corrections to the $Z$ coupling to right-handed fermions at the same
order.  We perform a global fit to the low energy data (see
\cite{Choudhury:2001hs} for details as to how the fit is
implemented) provides the $95\%$ C.L. $
q \langle \phi \rangle \gtrsim$ 2 TeV, 
whereas the bound on the
4-lepton contact-interaction from LEP-II \cite{Abbaneo:2001ix} is
$q \langle \phi \rangle \gtrsim$ $3.75$ TeV.

We therefore take the following example parameters:  
\begin{itemize}
\item{$\alpha_x\equiv g_x^2/4\pi = 1/35$ at a few TeV.  The beta-function 
coefficient for the gauge coupling $g_x$ is $b_x = 7 + 2 q^2$.  For the value 
$q=1/2$, the coupling runs semi-perturbatively at the GUT scale ({\it i.e.}, 
$\alpha_x(\Lambda_{GUT}) \sim 1$).}
\item{A $Z^\prime$ mass of 2.2 TeV (q=1/2), just above the
current LEP lower bound.}
\item{$m_\phi = 6.6$ TeV at low energies.  One loop corrections to the Higgs 
mass parameter from the supersymmetry breaking are finite and relatively small 
($<250$ GeV).  The two-loop RGE contribution from $m_\phi^2$ is smaller.}
\end{itemize}
The superpotential coupling $\lambda$ stays perturbative throughout the range 
of
scales for this choice of parameters (without fine-tuning the value of $B$).
The tree-level prediction for the Higgs mass 
can be computed from equation \ref{tree-higgs}.  In the decoupling limit with large $\tan\beta$,  
we find $m_{h^0}= 116$ GeV, consistent with the electroweak fits.  
The top-stop contribution to the one-loop effective potential results in an
actual Higgs mass which is larger than this value.  
Thus for most of the parameter space consistent with other
direct SUSY searches, the current direct search bound on the Higgs mass 
is satisfied.  

Unification of the standard three gauge couplings is still predicted to the 
percent level as
the new $U(1)_x$ coupling affects the running only at two loops.

We have presented a model which exemplifies a extra $D$-term 
contribution to the quartic potential of the Higgs.  
Since this section is intended as a warm-up, we have neither included the effects of kinetic mixing of this $U(1)$ with 
hypercharge, \footnote{We thank Hitoshi Murayama for reminding us of this point.} nor have we given details of how $D$-flatness is protected in the ultraviolet. Kinetic mixing will mix the $U(1)_x\  {\rm and\ } U(1)_Y$ $D$-terms and can be cancelled by a suitable counterterm.  If  $(m_{\phi}^2 - m_{\phi^c}^2)/\lambda \gtrsim 300$ GeV then the $U(1)_x$ $D$-term generates a tree-level mass for the Higgs field which introduces $\gtrsim 10 \%$ fine-tuning into the Electroweak VEV.  In the models below, neither of these points are relevant.

\section{An extra $SU(2)$}

In place of an extra $U(1)$, we now gauge an extra $SU(2)$ group.  The standard-model fields are charged 
under $SU(2)_1$ as the normal weak group and there is an additional group
$SU(2)_2$.  To break the $SU(2)_1 \times SU(2)_2$ to the diagonal 
subgroup we add an extra bi-doublet $\Sigma$ which transforms as a 
$(2, \overline{2})$.   Above the scale of diagonal symmetry breaking, 
the $SU(2)_1 \times SU(2)_2$ $D$-term is 
\begin{equation}
\frac{g_1^2}{8} \left( 
{\rm Tr} \left[ \Sigma^\dagger \sigma^a \Sigma \right] 
+ H^{\dagger}\sigma^a H -\overline{H} \sigma^a \overline{H}^{\dagger} + 
\ldots \right)^2  
+\frac{g_2^2}{8} \left( 
{\rm Tr} \left[ \Sigma \sigma^a \Sigma^\dagger \right] \right)^2 .
\label{eq:d-term}
\end{equation}
The superpotential $\mathcal{W}=\lambda  S \left( \frac{1}{2}\Sigma\Sigma + w^2 \right)$
with an additional soft-mass $m^2$ for $\Sigma$ leads to the scalar potential
\bea
  V_\Sigma & = & \frac{1}{2}B \Sigma \Sigma + h.c. + m^2 |\Sigma|^2 + 
\frac{\lambda^2}{4} |\Sigma \Sigma|^2.
\label{eq:potential}
\eea
Here, $\Sigma \Sigma$ is contracted with two epsilon tensors 
and $B=\lambda w^2$.
For suffiently large $B$, $\Sigma$ acquires a VEV,
$\langle \Sigma \rangle = u \mathbf{1}$, with 
$u^2=(B - m^2)/\lambda^2$,
which breaks $SU(2)_1 \times SU(2)_2$ to the diagonal subgroup. The minimum 
lies in a $D$-flat direction, leaving both Higgs fields massless. 

Under the remaining $SU(2)$, $\Sigma$ contains a complex triplet, $T$, 
along with a complex singlet.  Integrating out the real part of the heavy 
triplet at tree-level gives
the effective Higgs potential below the triplet mass,
\bea
&& \frac{g^2}{8} \:\Delta\: \left(H^{\dagger} 
\vec{\sigma} H- \overline{H} \vec{\sigma} \overline{H}^\dagger  \right)^2
+ \frac{g_Y^2}{8} \left( |\overline{H}|^2-|H|^2 \right)^2, 
 \nonumber \\
&&{\rm with\ \ }
\Delta=\frac{1+\frac{2 m^2}{ u^2}\frac{1}{g_2^2}}{1+\frac{2 m^2}{ u^2}
\frac{1}{g_1^2+g_2^2}}\hspace{.1in}  {\rm and \ \ } 
\frac{1}{g^2} = \frac{1}{g_1^2} + \frac{1}{g_2^2}.
\label{eq:delta}
\end{eqnarray}
The MSSM $D$-term is recovered in the limit $u^2 \gg m^2$ (no SUSY breaking),
for which SUSY protects the $D$-term below the gauge-breaking scale. 

As in the $U(1)$ case, electroweak symmetry breaking occurs under the same conditions as in the MSSM. 
 We find
the tree-level $W$ and $Z$ masses are corrected by the same relative
amount, $(1 - g^4 v^2 / 2 g_2^4 u^2 + ...)$ while the tree-level Higgs mass 
satisfies
\be
m_{h^o}^2 <  \frac{1}{2} \left( g^2 \Delta + g_Y^2 \right) v^2 
\cos^2{2 \beta} .
\ee
To maximize the upper bound, $\Delta$ should be made as 
large as possible by sending $g_1 \rightarrow \infty$,
$g_2 \rightarrow g$ and $m^2 \gg u^2$ by as much as possible without 
introducing fine-tuning. 

Precision electroweak constraints were analyzed in
\cite{Chivukula:2003wj} resulting in the $95\%$ C.L. constraint
$(1/2) (g/g_2)^4 (v/u)^2 \leq 2.1 \times 10^{-3}$. However, our setup
has an additional contribution to the oblique parameter $T$ due to a
small triplet VEV.  This results in a contribution $\Delta T \sim (4
\pi/ s_W^2 c_W^2) (g_1^4 / g^4) (M_W^2 u^2 / M_T^4)$, where $M_T$ is
the triplet mass.  The triplet VEV and larger $m_{h^0}$ partially compensate
each other in the $\Delta T$ piece of the electroweak fit.

A sample point in which 
perturbative unification is achieved with the right matter content at the GUT scale 
(see below for more details)
is $g_1(u) = 1.05$ and $g_2(u) = 0.83$. Precision electroweak
constraints and fine-tuning bounds are avoided for $m=2u=3.3$ TeV, which
implies $m_{Z'}=m_{W'}=2.5$ TeV. 
For this sample point, $\Delta = 2.3$ and $m_{h^0} = 129$ GeV at tree-level in the
large $\tan{\beta}$ and decoupling limits.  Again this Higgs mass, while large
enough to comfortably evade the LEP-II bounds, is consistent with the
electroweak fits.

The size of $g_1$ (and therefore $\Delta$) in 
the $SU(2)$ scenario was limited by its
large positive beta-function coefficient.  One can ameliorate this
situation by instead dividing the matter between the two $SU(2)$ groups such
that $g_1$ runs asymptotically-free and is thus larger at the weak scale---
leading to a larger value of $\Delta$.  We consider a non-universal 
model with the
Higgses and third family charged under $SU(2)_1$, while the first two
families are charged under $SU(2)_2$.

Yukawa couplings for the first two generations can be generated by adding
a massive Higgs-like pair of doublets $\overline{H}',H'$, that are charged under
$SU(2)_2$. They couple
to the first two generations via Yukawa-type
couplings and mix with the regular Higgses via superpotential operators
such as $\lambda' \overline{H} \Sigma H'$.  A supersymmetric mass
$\mu_{H'} > \langle\Sigma\rangle$ for the new doublets generates
naturally small Yukawa couplings
for the first two generations at low energies.

The constraints on a non-universal model, however, are more severe, as
there are tree-level non-oblique
corrections to the third family couplings \cite{Li:nk}.  We fit the
precision data, including the additional contribution to $\Delta T$
from the triplet VEV and find the ($95\%$ C.L.) constraint on $u$
as a function of $g/g_2$.  The strongest constraints occur for
$g^2/g_2^2 \rightarrow 0,1$.

We take the following example parameters:
\begin{itemize}
  \item $g_1(u) = 1.80, \ g_2(u) = .70$, inspired by a GUT
        with $g_1(\Lambda_{GUT}) = .97$. Additional spectator fields 
(see the full description at the end of the section for details) are included in the running to aid in unification.
  \item $u = 2.4$ TeV, above the lower limit from electroweak 
        constraints, giving $M_{W'}, M_{Z'} \sim 4.5$ TeV.
\item $m=10\  {\rm TeV}$. One-loop finite corrections to the Higgs 
mass parameter from supersymmetry breaking are $< 300$ GeV 
whereas two-loop RGE contributions can be somewhat
larger if one assumes high-scale supersymmetry breaking.
\end{itemize}
We find $\Delta =  6.97$ and $m_h =214$ GeV 
at tree-level in the large $\tan{\beta}$ and decoupling limits. 
Loop corrections to the effective potential from the 
top sector and the additional physics will make a 
relatively small shift in the tree-level result.  

Since $SU(2)_1$ is asymptotically free, we can push $\alpha_1(u)$ to
the perturbative limit, $\alpha_1(u) = 1$, by adjusting its high-scale
value.  Electroweak precision constraints for this $g/g_2$ require $u
\gtrsim 3.1$ TeV, while fine-tuning at the GUT-scale increases as we
tune the confinement scale and $u$ to coincide.  For $g_1(u) = 3.75$
and $g_2(u) = .66$, we choose $g_1(\Lambda_{GUT}) = 1.1$, tuned to be
within $1\%$ of its critical value. For $m=10$ TeV, we find $\Delta
\sim 20$ and $m_h \sim 350$ GeV in the large $\tan\beta$ and
decoupling limits. While the large Higgs mass gives a small positive
contribution to the S-parameter and a large negative contribution to the
T-parameter, this effect is offset in the global
fit by a positive T-parameter contribution from the small triplet VEV. Quantum
contributions to the Higgs mass parameter are of order $1.3$ TeV which
represents a fine-tuning of around 7\%.

One interesting feature of this model is that because there is a gauge coupling
larger than that of $SU(3)$ color, the top yukawa ``fixed point'' has a much larger value
than in the MSSM.  In this sense, a favorable region of parameter space includes some of $\tan\beta<1$
 which can both be consistent with the Higgs mass bound and avoid a Landau pole for the top Yukawa.

This model can also be made consistent with gauge coupling unification.
The full group $SU(3)_c \times SU(2)_1 \times SU(2)_2 \times U(1)_Y$
can be embedded in $SU(5)\times SU(5)$
\cite{Kribs:2002ew} broken by a bi-fundamental field at the GUT scale
with a vev $\langle\Xi\rangle = diag\{M,M,M,0,0\}$.  Gauge coupling unification
is predicted (with theoretical uncertainty beyond one-loop) because the 
standard model gauge couplings are only a function of the diagonal gauge
coupling.  At one loop, one can track the diagonal $SU(2)$ through its
beta-function coefficient $b$ as it is the sum of those of the two $SU(2)_i$.
It receives an extra $-6$ from the additional triplet of gauge bosons.  We include
two triplets charged under $SU(2)_2$ which, with the diagonal-breaking $\Sigma$ field,
contribute 
+6 to the diagonal beta function. We have also added an additional vector-like pair of
triplets to effectively complete a $5$ and $\overline{5}$ with the 
extra pair of Higgs-like fields (however, they should be from a split multiplet
as they must not share the Yukawa couplings with the doublets due to 
proton decay). With these additions, the $SU(2)$ model achieves
the same unification accuracy as in the MSSM at one loop.  Though there is a gauge coupling
that gets relatively strong, its two-loop effect is still small as $g_1$ is quite perturbative for
nearly all of the running.

\section{Conclusions and Outlook}
%\label{sec:conclusions}

The point of this paper is to show that  asymptotically-free gauge extensions of the MSSM 
can produce significant contributions to the Higgs quartic coupling --- and 
therefore the physical Higgs mass --- without destabilizing the weak scale.  
Breaking extra gauge groups in the multi-TeV range with a soft mass for the 
breaking field at the same scale leaves a non-decoupling contribution to the 
Higgs quartic potential.  Because of the $D$-term structure, there are no
$\log$-enhanced one-loop contributions to the Higgs soft mass and thus even 
after running from high-scales, the electroweak scale remains natural.

While technically natural, what could be the source of this higher
scale (few - 10 TeV) which is necessary for the extra gauge breaking
and the breaking-field's soft mass?  In fact, models of
anomaly-mediated supersymmetry breaking \cite{Randall:1998uk} and
gaugino-mediated supersymmetry breaking \cite{Kaplan:1999ac} provide
such a scale. Specifically, $\mu$-like terms are enhanced by a loop
factor in the former, while soft masses for bulk scalars are enhanced by
a volume factor in the latter.

In addition, there currently exists in the literature supersymmetric
models which make use of extended gauge sectors.  For example, extra
gauge groups are used to avoid negative squared masses for sleptons in
models of anomaly-mediated supersymmetry breaking
\cite{Arkani-Hamed:2000xj,Nelson:2002sa}.  It would be interesting to
calculate how much the quartic coupling can be enhanced in these models.  If
Higgs mass bounds increase, they may turn out to be more natural than
the MSSM.

An interesting question to ask is what happens if we allow the $SU(2)_1$ coupling
to blow up at the preferred breaking scale.  From arguments involving ``complementarity''
\cite{Fradkin:1978dv,Raby:1979my}, we speculate that the composite theory in the
infrared mimics the weakly coupled theory in the Higgs phase.  The Higgs and
the third generation would be composite and strongly coupled and therefore arbitrary
Higgs masses (consistent with unitarity bounds) would appear possible.  One remarkable
property of this model would be that gauge coupling unification would {\it still} only be
affected at the few percent level, as in \cite{Dimopoulos:2002bn}.  Of course it is
crucial that supersymmetry breaks at the same scale so a severe fine-tuning of scales
would be required.  It would be interesting to see to what extent that accident could have
a dynamical origin.

The natural regions of parameter space in these models leave behind 
extra gauge bosons with masses of order 2-5 TeV.  Due to the stronger gauge
coupling, these may be accessible at the LHC.  Thus, if superpartners
are discovered, a search for an extended gauge sector could be fruitful 
even if the Higgs mass is below 130 GeV.
\\ \\
{\it Note:}  Half a plenary talk at SUSY02 by L. Randall (who cited work in progress by N
. Arkani-Hamed, N. Weiner and herself) was devoted to an idea involving non-decoupling $D$-terms \cite{LisaTalk}. 

\acknowledgments
The authors have benefitted from discussions and general abuse from J.R. Espinosa
A. Nelson, M. Quiros, and C.E.M. Wagner,
and are grateful to the Aspen Center for Physics, at which much of
this work was envisioned.  A.D., D.E.K. and P.B. are 
supported by NSF Grants P420D3620414350 and P420D3620434350.
Fermilab is operated by Universities Research Association Inc.  under 
contract no. DE-AC02-76CH02000 with the DOE.

\end{document}